\documentclass{ws-procs9x6}

\begin{document}

\title{Physical processes behind the alignment effect}

\author{S. Mendoza \& J.~C. Hidalgo\footnote{\uppercase{C}urrent address: 
\uppercase{C}hurchill \uppercase{C}ollege, \uppercase{C}ambridge 
\uppercase{CB}3 \uppercase{ODS}, \uppercase{U}nited \uppercase{K}ingdom.} }

\address{Instituto de Astronom\'{\i}a, Universidad Nacional Aut\'onoma de
M\'exico, A.P.~70-264, Ciudad Universitaria, Distrito Federal CP 04510,
M\'exico\\ E-mail: sergio@astroscu.unam.mx, jch54@cam.ac.uk}


\maketitle

\abstracts{ The radio/optical alignment effect for small powerful radio
galaxies has been shown to be produced by shock waves formed by the
interaction of the head of the jet and/or cocoon with clouds embedded
in the interstellar/intergalactic medium.  We present here preliminary
results of analytical and numerical solutions that have been made to
account for the production of implosive shock waves induced by embedding
cold clouds in the radio lobe of expanding powerful radio sources.}

\section{Introduction}

 The radio/optical alignment effect in powerful radio galaxies was first
observed by Chambers and collaborators\cite{chambers87,mccarthy87}
and has been extensively investigated by Best and
collaborators\cite{best97,best98,inskip03,inskip03b}.  All these
observations show powerful radio galaxies that display enhanced
optical/UV continuum emission and extended emission line regions,
elongated and aligned with the radio jet axis. The expansion of
the radio source strongly affects the gas clouds in the surrounding
intergalactic/interstellar medium.

  Best and his collaborators\cite{best00} showed that the emission from
small radio sources (\( \lesssim 150 \, \textrm{pc} \)) were dominated
by shocks most probably generated by the interaction between the bow
shock of the jet and the surrounding medium.  For large radio sources
the emission appears to be the AGN itself.

  From the various theoretical models that have been proposed to account
for the emission observed, one of the most exotic ones was the idea that
the interaction of the jet with clouds embedded in the interstellar
medium induced the formation of stars.  Based on the original idea of
Begelman\cite{begelman89} we have constructed a model in which implosive
shock waves were driven in the clouds after finding themselves immersed
on very high pressure environments due to the passage of the jet.

\section{Implosive shock--waves} \label{implosive-shocks}

  When the head of an expanding extragalactic jet encounters clouds with
much smaller radii than the radius of the jet, then one can think that the
interaction hardly modifies the structure of the jet.  The collision
between the bow shock wave and/or the hot spot shock of the jet with 
clouds embedded in the interstellar and
intergalactic medium is quite complicated. A very general 2D description of
the collision between a plane parallel shock and a cloud was described in
detail\cite{klein94}. These calculations showed that after the passage of
the shock wave, shocks and rarefaction waves are formed\cite{mendoza00}
which lead to the destruction of the cloud.  

  At first approximation one can model the interaction of a shock
wave with a cloud as follows.  The clouds are originally in pressure
equilibrium with their surrounding interstellar or intergalactic medium.
After being swallowed by the expanding jet, the clouds find themselves on
an overpressured medium\cite{begelman89}.  If the post-shock pressure
is sufficiently large as compared to the pre-shocked pressure one
can guarantee the formation of an initial discontinuity that produces
an implosive shock wave and an expanding rarefaction wave, leaving a
tangential discontinuity which can now be identified as the border of
the cloud.

  Well known similarity solutions for the problem of a
non--relativistic explosive spherical shock waves have been found
in the past\cite{stanyukovich60}.  A similarity solution for the
problem of a relativistic shock wave was found by\cite{blandford76}.
Both approaches, relativistic and non--relativistic are based on the
idea that the energy content inside the shock wave is constant.

  Guderley\cite{guderley42}, Landau and
Stanyukovich\cite{stanyukovich60,daufm} found self--similar solutions
for the case of a spherical shock wave converging to a centre, the so
called implosive shock wave.   For the case of a relativistic implosive
shock wave we have found a similarity solution that generalises Landau
\& Staniukovich's model and takes into account some of the relativistic
ingredients introduced by Blandford and McKee\cite{blandford76} for the
relativistic explosive shock wave.  The complete mathematical description
of this model will be described elsewhere.  Here, we briefly mention the
most important results.  For this particular case, the Lorentz factor \(
\Gamma \) of the implosive shock wave is such that \( \Gamma^2 = A ( -t
)^{-m} \), where the constant \( m \) represents the similarity index.
Exactly as it happens for the non--relativistic implosive shock wave,
due to the fact that the energy is not conserved, one has to find
the similarity index by demanding non--singularities on the equations
of motion.  By doing that and assuming a polytropic index \( \kappa =
4/3 \) we found that \( m = 0.78460969 \).  With this similarity index it
was then possible to find numerical solutions for the density, pressure
and velocity profiles of the post--shocked material.

\section{Astrophysical consequences}

  With the model presented in Section \ref{implosive-shocks} it is
now possible to apply the results to typical clouds in the interstellar
medium.  Inside the jet of a typical FR--II radio galaxy, the velocity  of
the plasma is close to the velocity of light, and the equation of state of
the plasma is \( p = e / 3 \), where \( e \) is the proper energy density.
The pressure inside the jet has a typical value of \( 10^{-5} \textrm{Pa} \).
When this gas is shocked, the pressure grows to a very high value of \(
10^{-2} \textrm{Pa} \).  Under these circumstances, when a cold (\( \sim
10 \, \textrm{K} \)) dense (\( \sim 10^2 \textrm{cm}^{-3} \)) typical
interstellar cloud with characteristic radius of \( 1 \, \textrm{pc}
\), finds itself inside the radio lobe of a radio galaxy, then an
ultrarelativistic implosive shock wave is developed inside its structure.

  The post--shock temperature, immediately after the shock  reaches
values of \( 10^{13} \, \textrm{K} \).  The particle number density
at this point is \( 10^{4} \textrm{cm}^{-3} \).  The cooling time \(
\tau_\text{cool} \) of the post-shocked gas was calculated\cite{silk93} 
and its given by \( \tau_\text{cool} = 3.72 \times 10^{14}
\textrm{s} \).  In comparison, the collapse time of the implosive shock
wave is \( \tau_\textrm{coll} = 1.13 \times 10^{8} \textrm{s} \).
Since \( \tau_\textrm{cool} \gg \tau_\textrm{coll} \) the shock is
able to collapse completely and even reach a situation of bouncing
back as a kind of explosive shock wave.  At the time of collapse the
radius of the cloud diminishes to a value of \( 37600 \, \textrm{AU} \).
When this stage is reached, the post-shock values for the pressure, density
and temperature are found to be \( 0.01 \, \textrm{Pa} \), \( 1500 \,
\textrm{cm}^{-3} \) and \( 1.81 \times 10^{12} \textrm{K} \) respectively.
Under this circumstances, the squared ratio of the post--shocked Jeans
mass \( M_\text{J2} \) to the pre--shocked Jeans mass \( M_\text{J1} \)
of the cloud reaches a value of \( 6.6 \times 10^{34} \).  This means
that the Jeans mass of the cloud grows on a very great proportion and
so, a gravitational collapse can not occur at least for the adiabatic
solution presented here.  

  In summary, we have prooved that a gravitational collapse is not
possible under the model presented here.  So, under these circumstances no
star formation is induced by the passage of the radio jet through clouds
embedded in the interstellar medium of the host galaxy.  However, this
model suggests another form of shock radiation that might be presented
when clouds get embedded in the radio lobes of powerfull radio galaxies.

  It is important to note that radiation and self--gravity of the cloud was
not included on our calculations.  We intend to do a full 3D
simulation including these physical ingredients in the future.

\section*{Acknowledgements}
  S Mendoza thanks support granted by CONACyT (41443).  JC Hidalgo 
acknowledges the support granted by Fundaci\'on UNAM. 


\end{document}